\documentclass[aps,pra,reprint,groupedaddress,longbibliography]{revtex4-2}
\usepackage{graphicx}
\usepackage{verbatim}
\usepackage{subfigure}
\usepackage[english]{babel}
\usepackage{siunitx}
\usepackage{pifont}
\usepackage{amsmath}
\usepackage{adjustbox}
\usepackage{wasysym}
\usepackage{mathtools} 
\usepackage{upgreek}
\usepackage{multirow}
\usepackage{xcolor}
\usepackage{graphicx}
\usepackage{parskip}
\usepackage[margin=0.75in]{geometry}
\begin{document}

\title{Combining laser cooling and Zeeman deceleration for precision spectroscopy in supersonic beams  } 
\author{Gloria Clausen$^1$, Laura Gabriel$^2$, Josef A. Agner$^1$, Hansj\"urg Schmutz$^1$, Tobias Thiele$^4$, Andreas Wallraff$^{3,4}$, Fr\'ed\'eric Merkt$^{1,3,4}$}
\affiliation{$^1$Department of Chemistry and Applied Biosciences, ETH Zurich, CH-8093 Zurich, Switzerland\\
$^2$Department of Civil, Environmental and Geomatic Engineering, ETH Zurich,CH-8093 Zurich, Switzerland \\
	$^3$Quantum Center, ETH Zurich, CH-8093 Zurich, Switzerland\\
	$^4$Department of Physics, ETH Zurich, CH-8093 Zurich, Switzerland}

\date{\today}

\begin{abstract}
Precision spectroscopic measurements in atoms and molecules play an increasingly important role in chemistry and physics, e.g., to characterize structure and dynamics at long timescales, to determine physical constants, or to search for physics beyond the standard model of particle physics. In this article, we demonstrate the combination of Zeeman deceleration and transverse laser cooling to generate slow (mean velocity of 175 m/s) and transversely ultracold ($T_\perp \approx 135 \,\mu$K) supersonic beams of metastable $(1s)(2s)\,^3S_1$ He (He$^*$) for precision spectroscopy. The curved-wavefront laser-cooling approach is used to achieve large capture velocities and high He$^*$ number densities. The beam properties are characterized by imaging, time-of-flight and high-resolution spectroscopic methods, and the factors limiting the Doppler widths in single-photon spectroscopic measurements of the $(1 s)(40 p) \,^3 P_J \, \leftarrow (1 s)(2 s) \,^3 S_1$ transition at UV frequencies around $1.15\times 10^{15}$ Hz are analyzed. In particular, the use of skimmers to geometrically confine the beam in the transverse directions is examined and shown to not always lead to a reduction of the Doppler width. Linewidths as narrow as 5 MHz could be obtained, enabling the determination of line centers with a precision of $\Delta \nu/\nu$ of $4\times 10^{-11}$ limited by the signal-to-noise ratio. Numerical particle-trajectory simulations are used to interpret the experimental observations and validate the conclusions.
\end{abstract}

\maketitle

\section{Introduction}
In recent years, a lot of effort has been invested in developing table-top precision-spectroscopy experiments in the optical-frequency range involving cold and ultracold samples of atoms and molecules. The experiments exploit the linewidth reduction resulting from the narrow velocity distributions and long interaction times of the samples with the radiation field. Their scientific goals include the determination of physical constants such as the fine-structure constant (see, e.g., Ref. \onlinecite{hudson06a}), the Rydberg constant and nuclear charge radii (see, e.g., Refs. \onlinecite{ beyer17b,grinin20a,scheidegger24a}),  the test of fundamental theories of atomic and molecular structure in  ``calculable" few-body systems \citep{hoelsch19a,alighanbari20a,patra20a}, and searches for physics beyond the standard model of particle physics \citep{
hudson11a, andreev18a,delaunay23a, roussy23a}. An attractive scheme for these experiments consists in generating highly collimated slow beams of the atoms or molecules of interest and probing transitions of narrow natural widths in field-free space. \\ \indent Slow beams of numerous atoms and molecules can be generated by hydrodynamic expansions from cold dense buffer-gas-cooled samples \citep{messer84a,maxwell05a, hutzler12a}, by laser Zeeman-slowing schemes \citep{phillips82a, kaebert21a}, and by multistage Stark \citep{bethlem99a,osterwalder10a}, Rydberg-Stark \citep{procter03a,vliegen04a} and Zeeman \citep{vanhaecke07a, hogan08d,narevicius08a} deceleration of supersonic beams.
When recording single-photon transitions, beams with narrow transverse-velocity distributions are required to minimize line broadening by the first-order Doppler effect. Such beams can be formed either by restricting the gas-expansion cones with skimmers - however, at the cost of reduced particle density and signal strength- or by one dimensional (1D) transverse laser cooling along the propagation direction of the spectroscopy laser. The latter approach is particularly attractive because efficient laser-cooling schemes exist for numerous atoms and an increasing number of molecules \citep{fitch21a,chae23a} including not only diatomic molecules such as SrF \citep{deMille13a}, CaF \citep{zhelyazkova14a}, YbF \citep{lim18a}, BaF \citep{zhang22a}, BaH \citep{mcnally20a}, CaH \citep{vazquez22a} and AlF \citep{hofsäss21a} but also polyatomic molecules such as CaOH \citep{vilas22a}, YbOH \citep{augenbraun20a}, SrOH \citep{kozyryev17a} and CaOCH$_3$ \citep{mitra20a}. \\ \indent Most of these molecules have a permanent electric dipole moment and are paramagnetic.
Consequently, combining multistage Zeeman deceleration methods to generate slow beams with 1D transverse laser cooling represents a very attractive route for precision measurements in atoms and molecules. This was immediately noted \citep{tarbutt13a, vandenberg12a, vandenberg14a, wall16a} following the initial reports of 1D molecular laser cooling \citep{shuman10a, barry12a, barry12b}. Indeed, transverse laser cooling of atoms or molecules starting from narrow and slow beams requires fewer absorption-emission cycles than the full three-dimensional laser cooling of a thermal gas.  \\ \indent We report here on the combination of multistage Zeeman deceleration with transverse cooling using the curved-wavefront approach \citep{rooijakkers96a, aspect90a,pereira_dos_santos01a, keller14a,  chen20a} for the purpose of precision measurements of transitions from the $(1s)(2s)\,^3S_1$ metastable state of He (He$^*$ hereafter) to high $np$ Rydberg states. The advantage of this approach is its large acceptance velocity, which is sufficient to cool all atoms exiting a Zeeman decelerator. This aspect is crucial to achieve a high signal-to-noise ratio in spectroscopic measurements. The emphasis of our investigation is placed on the characterization and optimization of the spatial and velocity distributions using beam imaging and measurements of time-of-flight distributions and Doppler profiles. 
\section{Experimental Setup and methods}
\subsection{Overview}
\begin{figure}[h]
\includegraphics[trim=0cm 0cm 0cm 0cm, clip=true, width=1.0\columnwidth]{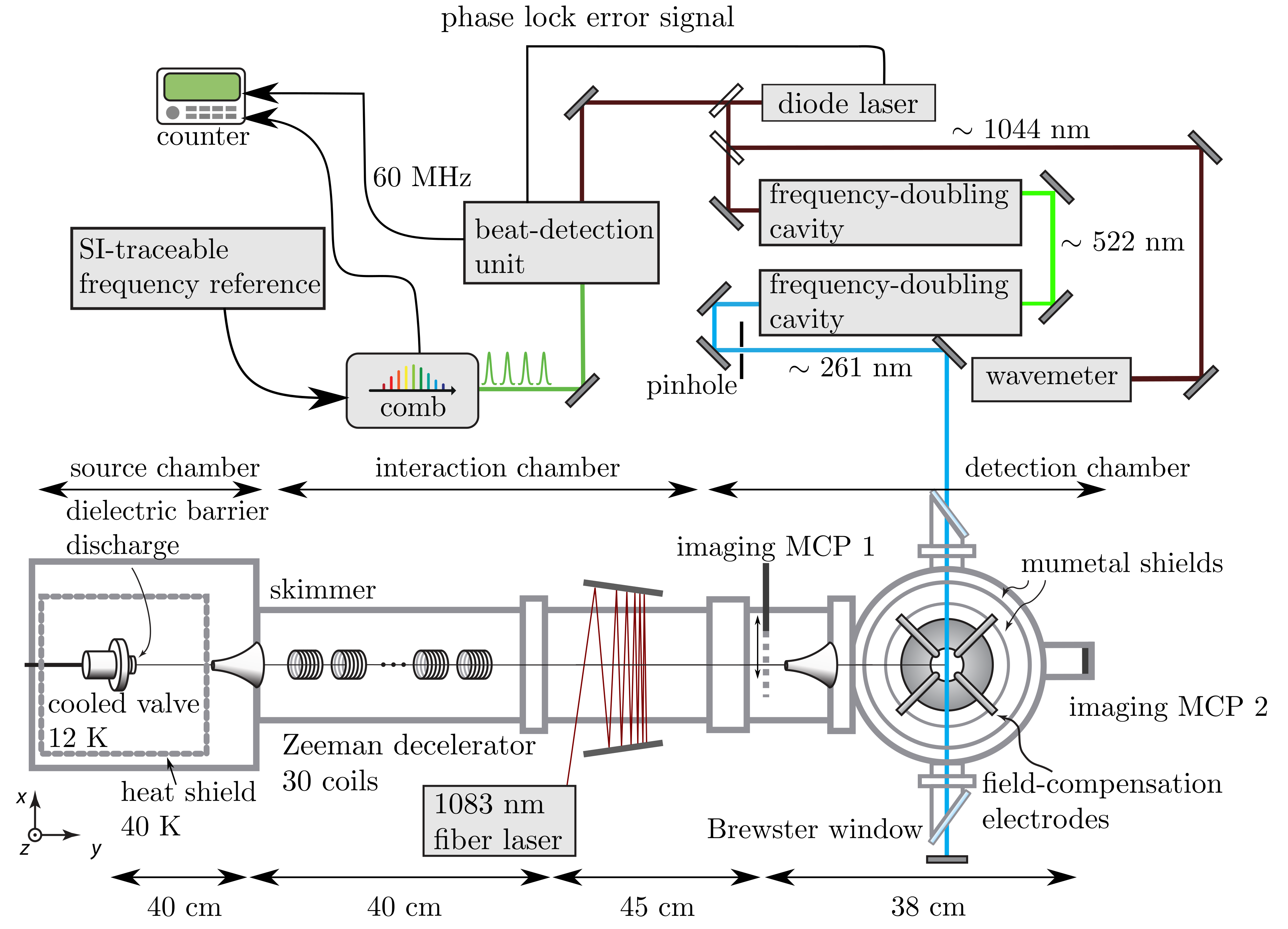}
\caption{\label{fig:1}Schematic diagram of the experimental setup. Top: Laser system including the optical frequency comb used for frequency calibration. Bottom left: Valve and discharge electrodes used to generate the supersonic expansion of metastable helium. Bottom center: Vacuum chamber, with Zeeman decelerator and transverse laser cooling for atomic-beam manipulation and two imaging microchannel plate (MCP 1 and 2) detectors for beam diagnostics. Bottom right: Photoexcitation chamber, where the atomic beam and the photoexcitation-laser beam cross at near-right angles within a double layer of magnetic  shielding and electric-stray-field-compensation electrodes.}
\end{figure}

The experimental setup is shown schematically in
Fig. \ref{fig:1}. Several of its main components have been described previously \citep{clausen21a, clausen23a}. The experiment is conducted using a supersonic-beam apparatus under high-vacuum conditions. A supersonic beam of He$^*$ is produced in the source chamber using a pulsed valve. In the interaction chamber, a multistage Zeeman decelerator and a transverse-laser-cooling section are used to slow down and cool the He$^*$ atoms in the beam. Their spatial and velocity distributions are then determined in the detection chamber. This chamber includes a photoexcitation region, in which the Doppler profiles of the $(1 s)(n p) \,^3 P_J \, \leftarrow (1 s)(2 s) \,^3 S_1$ transitions are measured, and two on-axis imaging microchannel plate (MCP) detectors, with which the transverse- and longitudinal-velocity distributions are monitored.
\subsection{Supersonic beam}
The source chamber consists of a thermal shield cooled to 40 K by the first stage of a pulse-tube cryocooler and a pulsed valve thermalized at 12 K with the second cooling stage.
A supersonic
beam of helium atoms with a mean velocity around 480 m/s
is emitted into vacuum from the pulsed valve at a repetition
rate of 8.33 Hz and pulse lengths of 20 $\mu$s. A dielectric-barrier discharge
at the nozzle orifice excites ground-state He to He$^*$ and the metastable atoms are entrained in the supersonic expansion. A skimmer with a 1-cm-diameter orifice located 40 cm downstream from the nozzle orifice limits the beam-expansion angle at the entrance of a Zeeman decelerator described in Sec. \ref{sec:zeeman}.
\subsection{Laser Systems}
\label{sec:laser_setup}
\begin{figure}[h]
\includegraphics[trim=0cm 0cm 0cm 0cm, clip=true, width=0.95\columnwidth]{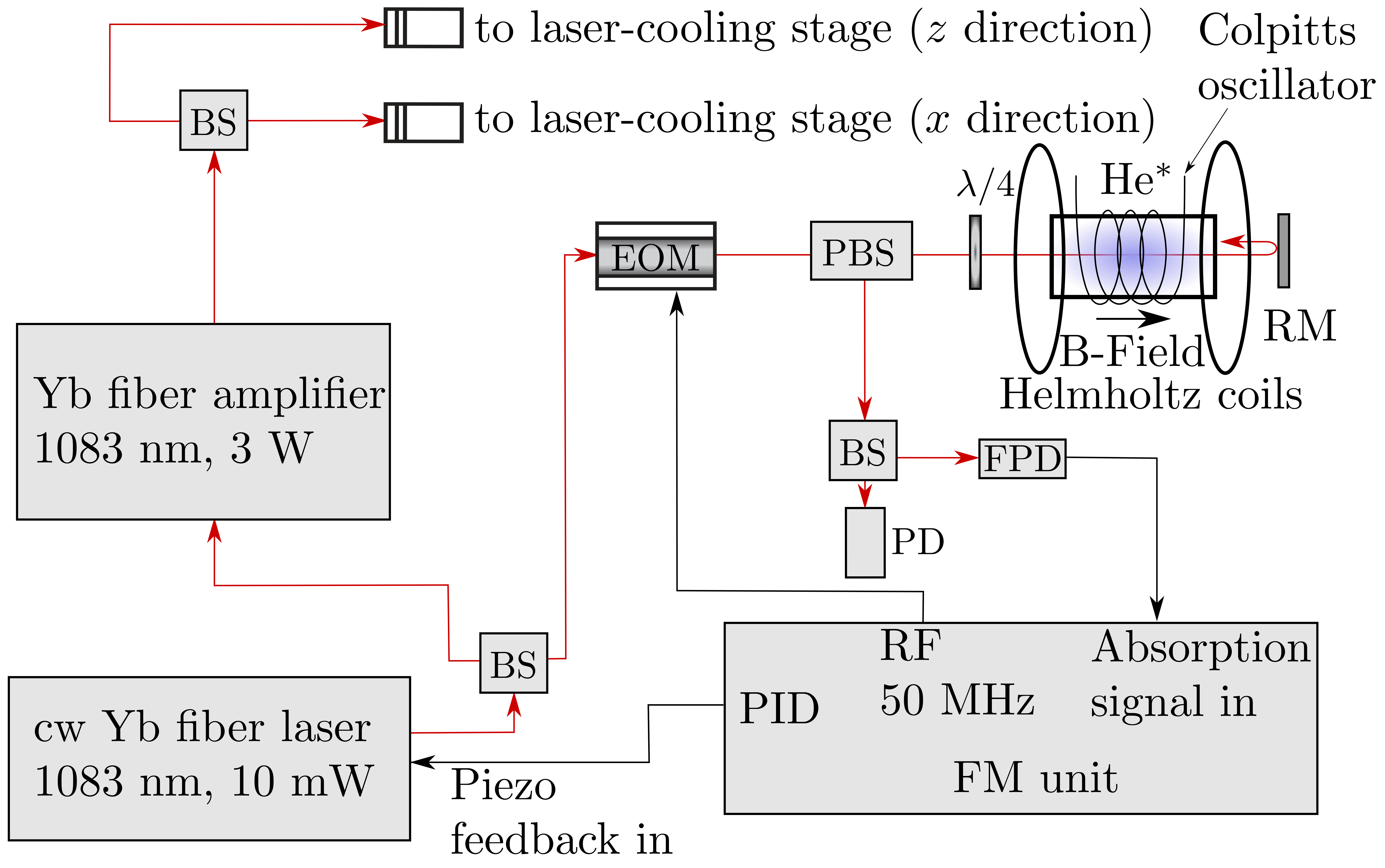}
\caption{\label{fig:2}Schematic diagram of the laser system used for transverse laser cooling of a supersonic beam of He$^*$. The laser beams are indicated as red lines and the electronic connections are indicated as black lines. (BS: 50:50 beam splitters; PBS: polarizing beam splitters; EOM: electro-optical modulator; FM unit: frequency-modulation locking unit; FPD: fast photodiode; PD: photodiode; RF: radiofrequency; PID controller: proportional-integral-derivative controller; RM: reflective mirror). See text for details. }
\end{figure}

Two laser systems are used in the experiments, one for the transverse laser cooling of He$^*$ and the other for the precision measurements of Rydberg states in triplet He. \\ \indent
 The main components of the laser system used for transverse  cooling of He$^*$ using the $2\,^3P_2\,\leftarrow\,2\,^3S_1$ closed cycle at 1083 nm are displayed schematically in Fig. \ref{fig:2}. The 1083 nm laser radiation is generated from a continuous-wave (cw) ytterbium fiber laser releasing 10 mW of output power. Half of this power is used for locking the laser frequency to the $2\,^3P_2 \,\leftarrow 2\,^3S_1$ transition using saturated-absorption spectroscopy \citep{preston96a} of He$^*$ in a discharge cell surrounded by a pair of coils in Helmholtz configuration (see Fig. \ref{fig:2}). Inside the cell, He$^*$ is generated in a radio-frequency (rf) discharge using a Colpitts oscillator \cite{moron12a}. A frequency-modulation locking scheme is employed to keep the laser on resonance with the $2\,^3P_2\,\leftarrow\,2\,^3S_1$ transition \citep{bjorklund80a, bjorklund83a}. The laser frequency is modulated at 50 MHz using an electro-optical modulator (EOM). The saturated-absorption signal is measured with a fast photodiode (FPD) and is demodulated with the 50-MHz rf signal to generate an error signal. A proportional-integral-derivative (PID) controller converts the error signal into a feedback signal, which is fed to the piezo voltage controller of the fiber laser. To tune the laser frequency, the $2\,^3P_2 \,\leftarrow 2\,^3S_1$ transition frequency of the He$^*$ atoms in the cell is adjusted by varying the magnetic field generated by the current flowing through the Helmholtz coils. The second half of the output power of the fiber laser is amplified with an ytterbium-doped fiber amplifier to produce 1--3 W of laser power. A beam splitter is used to generate the two beams of equal intensities required for transverse cooling of the He$^*$ beam in the horizontal ($x$) and vertical ($z$) directions (see Fig. \ref{fig:1}). The laser beam diameter in the laser-cooling region is about 5 mm. 
\\  \indent
The single-mode cw laser radiation at $\sim260$ nm with a frequency $\nu_\mathrm{L}$ used to record spectra of He $np$ Rydberg states from He$^*$ is generated by frequency quadrupling the $\sim1040$ nm output of a tapered-amplifier diode laser in two successive doubling cavities equipped with  nonlinear frequency-upconversion crystals. The laser is phase locked to an ultra-stable optical frequency comb and has a linewidth [full width at half maximum (FWHM)] of $\leq 80$ kHz at the fundamental frequency (see Ref.  \onlinecite{clausen23a} for details). The beam diameter in the photoexcitation region is about 1 mm.
\subsection{Multistage Zeeman Decelerator}
\label{sec:zeeman}
A multistage Zeeman decelerator uses an array of solenoids to generate inhomogeneous magnetic-field pulses with which  beams of paramagnetic atoms or molecules are decelerated \citep{vanhaecke07a}. The decelerator design and operation principles have been described earlier in Refs. \onlinecite{jansen20a, semeria18a,jansen16a, motsch14a}. The array consists of 30 coils with 60 windings each (length of 7 mm, inner radius $r_\mathrm{coil}$ of 3.5 mm) through which currents up to 230 A are pulsed, resulting in magnetic fields of $\sim 2$ T on axis. To achieve a phase-stable deceleration, the current pulses through the coils are switched on successively before the atomic cloud arrives at the entrance of a specific coil and switched off before the cloud has passed the center of that coil (see Ref. \citep{wiederkehr11a}). The timings for the sequential switching are calculated for a synchronous particle representing an ideal atomic beam pulse with a given starting velocity, taking into account the $8.5 \,\mu$s rise and fall times of the magnetic-field pulses (see Ref. \cite{wiederkehr11a} for details). The switch-off time of the coil currents is defined relative to the position of the synchronous particle inside the coil and is expressed as a phase angle. A phase angle of $90^\circ$ corresponds to a switch-off position at the center of the coil and removes the largest amount of kinetic energy per coil. The deceleration sequence is optimized for an initial forward velocity of the synchronous particle of 480 m/s, and a phase angle of $35^\circ$ enables the phase-stable deceleration of He$^*$ in low-field-seeking states ($M_S=-1$) to about 150 m/s when using all 30 coils. Larger final velocities of the decelerated beam are obtained by reducing the number of coils through which currents are pulsed. 
\begin{figure}[h]
\includegraphics[trim=0cm 0cm 0cm 0cm, clip=true, width=0.9\columnwidth]{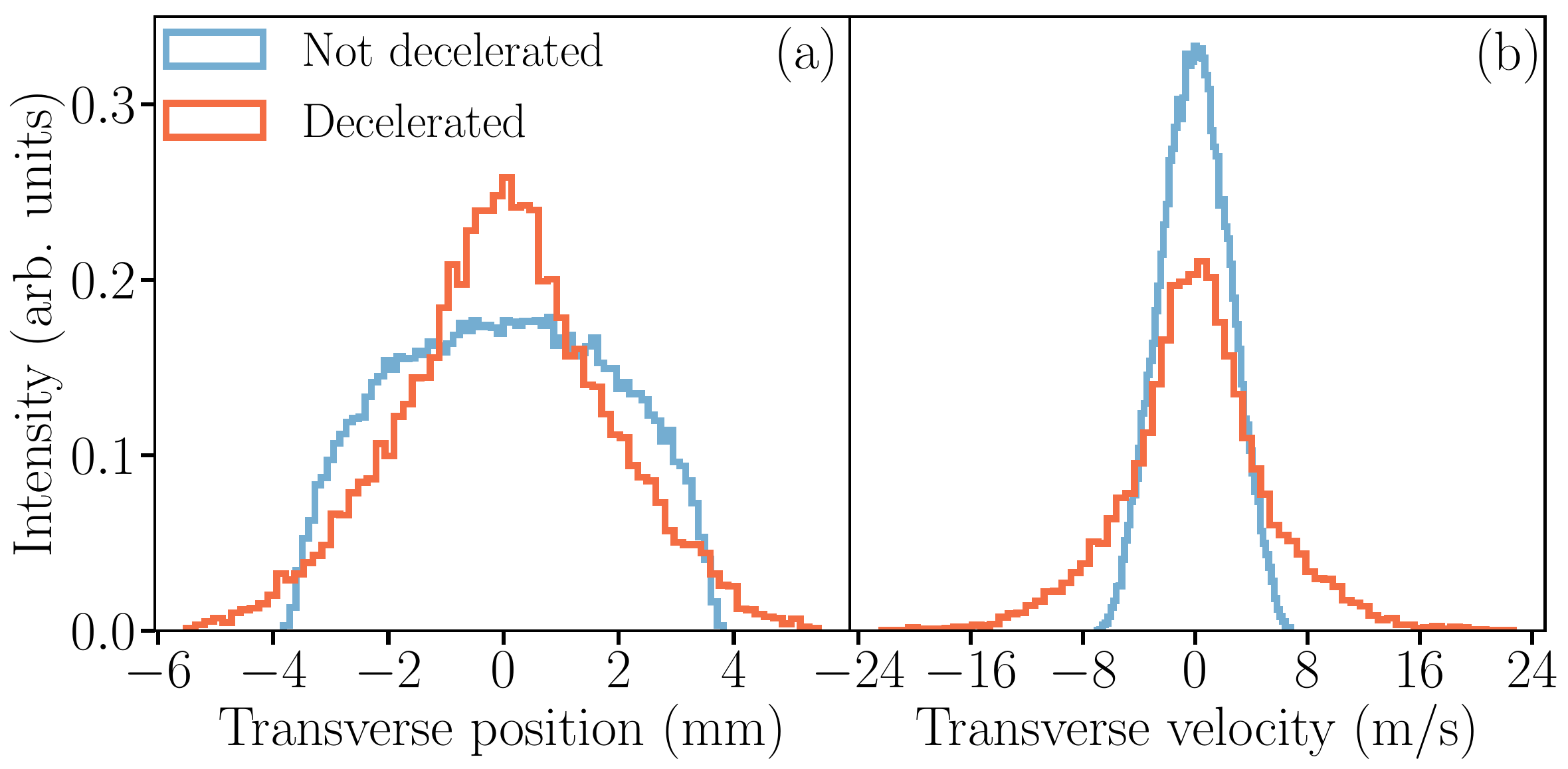}
\caption{\label{fig:3} Particle-trajectory simulations of the (a) transverse-position and (b) transverse-velocity distributions of the He$^*$ metastable beam at the exit of the Zeeman decelerator in the absence of deceleration (light blue) and for the decelerated samples with final velocities of 175 m/s (orange). See text for details.  }
\end{figure}
\\ \indent Figure \ref{fig:3} shows the predicted transverse-position (a) and transverse-velocity (b) distributions of a He$^*$ sample decelerated to a final velocity of 175 m/s (orange traces) and of the nondecelerated atomic beam (light-blue traces) at the end of the decelerator   obtained in Monte-Carlo particle-trajectory simulations \citep{wiederkehr11a}. \\ \indent In the absence of deceleration, the expansion angle of the He$^*$ beam directly behind the Zeeman decelerator is given by \begin{equation}
\gamma_1 =\arctan\frac{r_\mathrm{coil}}{R} \approx 4.4\,\mathrm{mrad}, \end{equation} where $r_\mathrm{coil}$ is the inner radius of the coils and $R$ is the distance between the valve orifice and the end of the decelerator coil array (80 cm, see Fig. \ref{fig:1}). With these parameters, the transverse-position spread of the undecelerated atomic beam at the end of the coil array is limited to the interval $[-3.5\,\mathrm{mm},\,3.5\,\mathrm{mm}]$ and its transverse-velocity spread is limited to the interval [$-6$ m/s, 6 m/s]. The velocity spread is larger than the spread [$-2$ m/s, 2 m/s] one would obtain from the geometric expansion of a point source at the valve orifice because of the extended volume of the discharge generating He$^*$ (see Sec. \ref{sec:results_cooling} and Fig. \ref{fig:6}(e) for additional information). When decelerating the atomic beam, the transverse-velocity spread increases by the focusing/defocusing effects of the magnetic fields on the He$^*$ beam. These lead to broadened wings in the distribution and to a half width at half maximum of the distribution of $\pm7.5$ m/s compared to $\pm3.8$ m/s for the nondecelerated beam (see Fig. \ref{fig:3} (b)). The broader distribution of transverse velocities results in a rapid decrease of the He$^*$ density along the beam propagation axis and to a large reduction of the signal-to-noise ratio of the spectra recorded in the detection region. 
  \subsection{Laser cooling } 
The laser-cooling section is located immediately after the Zeeman decelerator and is used to narrow down the transverse-velocity distribution of the atomic beam and minimize the corresponding Doppler broadening and to increase the number of atoms in the spectroscopy region to improve the signal-to-noise ratio. Transverse laser cooling of the He$^*$ beam is achieved using the closed-cycle transition $2\,^3P_2 \,\leftarrow 2\,^3S_1$ at 1083 nm \citep{chang14a}. The transition has a low saturation intensity of 0.167 mW/cm$^2$ and efficiently cools the light helium atoms \citep{zheng19a}. For the narrow transverse-velocity distributions of the He$^*$ beam at the end of the decelerator, $\sim60$ optical cycles are needed to reach the Doppler-cooling limit under ideal conditions.
\begin{figure}[h]
\includegraphics[trim=0cm 0cm 0cm 0cm, clip=true, width=0.8\columnwidth]{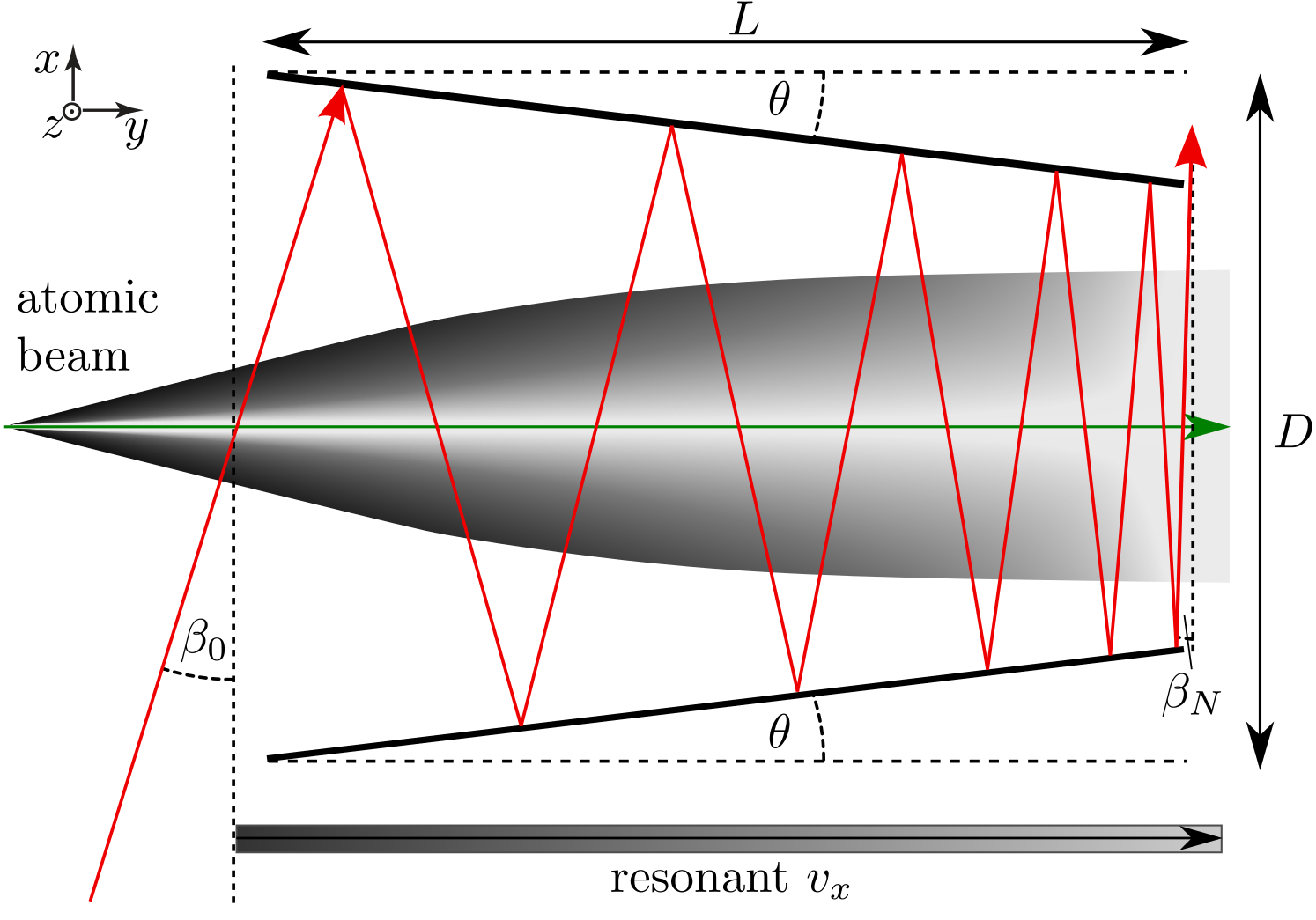}
\caption{\label{fig:4}Schematic diagram of the propagation of the cooling-laser beam (red lines) within a pair of tilted mirrors (black solid lines) illustrating the principle of the curved-wavefront laser-cooling approach \citep{rooijakkers96a, aspect90a,pereira_dos_santos01a, keller14a,  chen20a}. The incident laser beam makes an angle $\beta_0$ with respect to the normal axis (dashed vertical line) and leaves the laser-cooling region at an angle $\beta_N$. The mirrors are tilted with an angle $\theta$ with respect to the He$^*$-beam propagation axis (dashed horizontal line). The atomic-beam expansion is indicated by the grey cone and the transverse-velocity distribution along the $x$ axis is depicted by the different shades of grey.} \end{figure}
\\ \indent We employ the curved-wavefront laser-cooling approach \citep{rooijakkers96a, aspect90a,pereira_dos_santos01a, keller14a,  chen20a} for each of the two orthogonal transverse directions (horizontal ($x$) and vertical ($z$)), as schematically illustrated in Fig. \ref{fig:4} for the $x$ direction. In this approach, the cooling laser is reflected multiple times by two mirrors that are slightly tilted with respect to the propagation axis of the supersonic beam. The pointing of the wavevector of the laser follows a curved line, which enables one to match the laser frequency to the Doppler-shifted resonance frequency as the atoms propagate through the cooling region. The main advantage of this approach over commonly used two-dimensional molasses for cooling supersonic expansions is the broad transverse-velocity acceptance \citep{lett89a}. In each direction, the cooling-laser beam is reflected $N$ times between a pair of mirrors with surfaces tilted by angles of $\pm \theta$ with respect to the He$^*$ propagation ($y$) axis. The tilt angle between the mirrors changes the intersection angle between laser and atomic beams, and thus the Doppler shift, at each reflection (see Fig. \ref{fig:4}). The laser frequency is kept fixed at the Doppler-free $2\,^3P_2\,\leftarrow\,2\,^3S_1$ transition frequency, so that atoms flying perpendicularly to the laser beam are cooled most efficiently. The incidence angle of the laser beam $\beta_0$ (defined in Fig. \ref{fig:4}) determines the maximum transverse capture velocity $v_{\perp ,\mathrm{ max}}$ because it corresponds to the largest deviation angle from the normal axes of the atomic beam ($x$ and $z$ axes) and thus to the largest Doppler shift. As the atoms move forward, the angle between the laser beam and the $x$ ($z$) axis decreases and the resonance condition is fulfilled for atoms with progressively smaller transverse velocities $v_\perp$ (see grey scale at the bottom of Fig. \ref{fig:4}). \\ \indent The tilt angle $\theta$ of the mirrors determines the final intersection angle $\beta_N$ between the $N$-th reflection of the laser beam and the atomic beam and thereby the final transverse velocity $v_{\perp,\mathrm{min}}$ of the atoms at the end of the laser-cooling section. At each reflection, the incidence angle is reduced by $2\theta$, resulting in an angle $\beta_n$ after the $n$-th reflection given by $\beta_n=\beta_0-2n\theta$ \cite{ritterbusch09a,welte10a}.
For a given length $L$ and a given separation $D$ of the mirror-pair setup (see Fig. \ref{fig:4}), the incidence angle $\beta_0$ and the opening angle $\theta$ must be carefully adjusted to maximize the total number $N$ of reflections while guaranteeing that the laser beam leaves the two-mirror system at the end of the transverse-cooling stage. In particular, the condition $\theta_\mathrm{max}\leq\frac{D}{4}\frac{\beta_0^2}{L}$ must be fulfilled (see Ref. \citep{ritterbusch09a} for a complete derivation). \\ \indent Geometrical constraints in our experimental setup and the finite diameter of the laser beam ($\sim 5$ mm) limit the incident angle $\beta_0$ to values above $0.05\,$rad, corresponding to an absolute transverse velocity $|v_{\perp, \mathrm{ max}}|= 24$ m/s for the He$^*$ atoms that can still be efficiently cooled. The mirrors have a length $L$ of $15$ cm and are separated by a distance $D$ of $25$ cm, which results in $\theta_\mathrm{max}\approx 1$ mrad for the minimal $\beta_0$ value of 0.05 rad. The maximum number of reflections $N$ is given by $N=\frac{\beta_0}{2\theta_\mathrm{max}}\approx 25$ in our transverse-laser-cooling setup for $\beta_0=0.05$ rad. 
\subsection{Measurement of the velocity distribution of the He$^*$ beam} \label{sec:measurement}The transverse- and longitudinal-velocity distributions are measured using three separate detection schemes. The first scheme uses two slide-in on-axis imaging MCP assemblies, each consisting of a phosphor screen and a CCD camera, separated by 38 cm, the first one (MCP1) being positioned 40 cm downstream from the laser cooling section (see Fig. \ref{fig:1}). The transverse-velocity distribution can be extrapolated from the images of the spatial distributions monitored at the two imaging MCP detectors because the atomic beam expands linearly in the field-free region between the two MCPs. For the measurements of the transverse-velocity distribution with these two imaging MCP detectors, all components (electrode stack, skimmers, mumetal shields) that could obstruct the field of view were removed. \\ \indent
In the second scheme, the transverse-velocity distribution of the atomic beam is derived from the Doppler profile of individual transitions of He$^*$ to $np$ Rydberg states. About 60 cm beyond the laser-cooling section, the atomic beam is crossed by a single-mode laser beam ($\lambda \approx 260$ nm, see Sec. \ref{sec:laser_setup}) at near-right angles on the axis of a double-layer mumetal shield. The He$^*$ atoms are photoexcited to $(1s)(n p) \,^3 P_J$ Rydberg states. The principal quantum number $n$ is typically chosen around 40 because the broadening resulting from the spin-orbit splitting between the three fine-structure components $J=1,2,3$ of the $np$ Rydberg state scales as $n^{-3}$ and is negligible in comparison to the expected linewidth \citep{drake99a}. At the same time, the Stark shifts and line broadening from residual stray-electric fields in our experimental setup are negligible at $n=40$, as demonstrated previously \cite{clausen21a}. \\ \indent Pulsed-field ionization (PFI) is used to ionize the Rydberg atoms and the resulting ions are extracted towards an MCP detector (not shown in Fig. \ref{fig:1}) in the direction perpendicular to the laser and atomic beams. The first-order Doppler shift $\Delta \nu_\mathrm{D}$ of the transition frequency of an atom with a transverse velocity $v_\perp$ along the laser-propagation axis is given by $\Delta \nu_\mathrm{D}(v_\perp) = \nu(v_\perp)-\nu_0= \frac{v_\perp}{c}\nu_\mathrm{L}$, which allows the reconstruction of the transverse-velocity distribution from the spectral line shape of an individual transition. 
\\ \indent
In the third scheme, the longitudinal-velocity distribution of the He$^*$ beam is extracted by monitoring the He$^+$-ion signal generated by PFI following resonant laser photoexcitation to $np$ Rydberg states as a function of the delay between the valve opening and the PFI pulse. The resulting distribution, equivalent to a time-of-flight spectrum, is converted into a longitudinal-velocity distribution using the known distances between the valve orifice, the Zeeman decelerator and the photoexcitation spot.
\section{Results}
\subsection{Laser cooling}
\label{sec:results_cooling}
\begin{figure*}[]\includegraphics[width=0.9\textwidth]{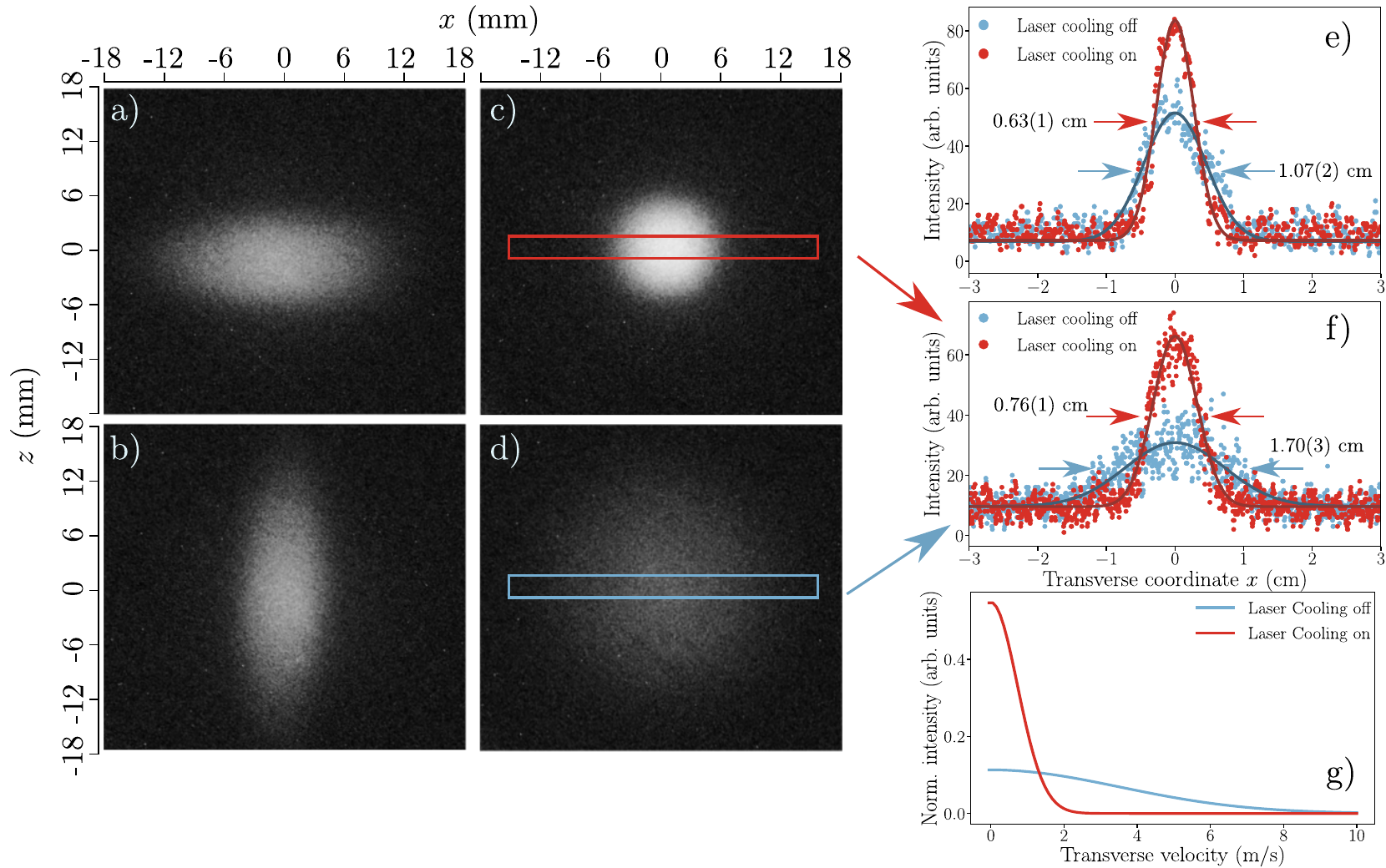}
\caption{\label{fig:5} (a)-(d) Images of the atomic He$^*$ cloud recorded with MCP 2 (see Fig. \ref{fig:1}) illustrating the effect of transverse laser cooling. (a) Cooling in the vertical ($z$) direction only. (b) Cooling in the horizontal ($x)$ direction only. (c) Cooling in both $x$ and $z$ directions. For comparison, (d) shows the image of the atomic beam recorded without laser cooling. (e) and (f): Atom-density profiles along the center of the atomic cloud recorded in the $x$ direction  with the imaging MCP 1 (e) and imaging MCP 2 (f), respectively. The red and light-blue profiles correspond to measurements with and without laser cooling. (g) Corresponding normalized transverse-velocity distributions along the $x$ directions extracted from these images (see text for details).    }
\end{figure*}
\begin{figure}[]
\hspace{1 mm}\includegraphics[width=0.83\columnwidth]{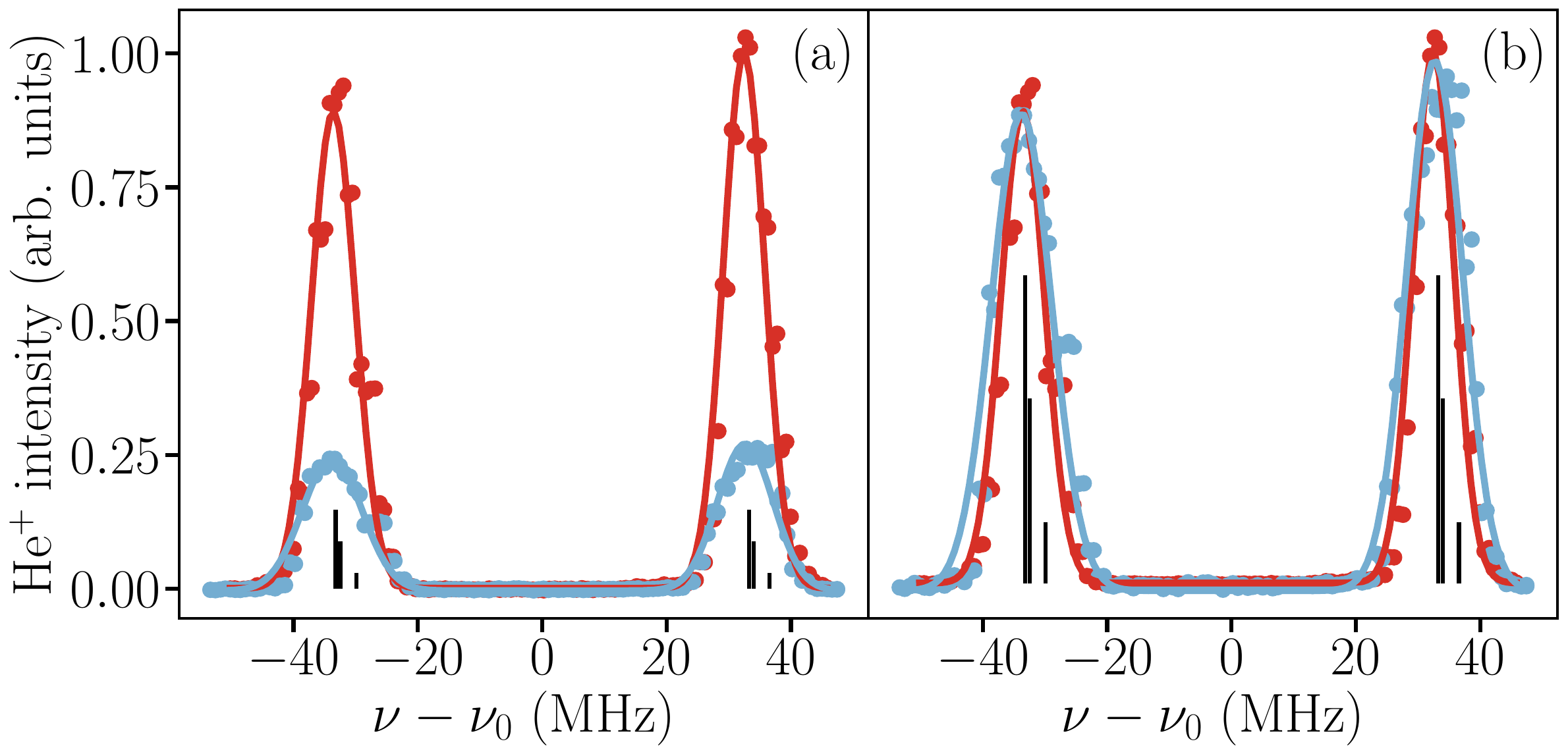}\\ \hspace{3 mm}\includegraphics[width=0.83\columnwidth]{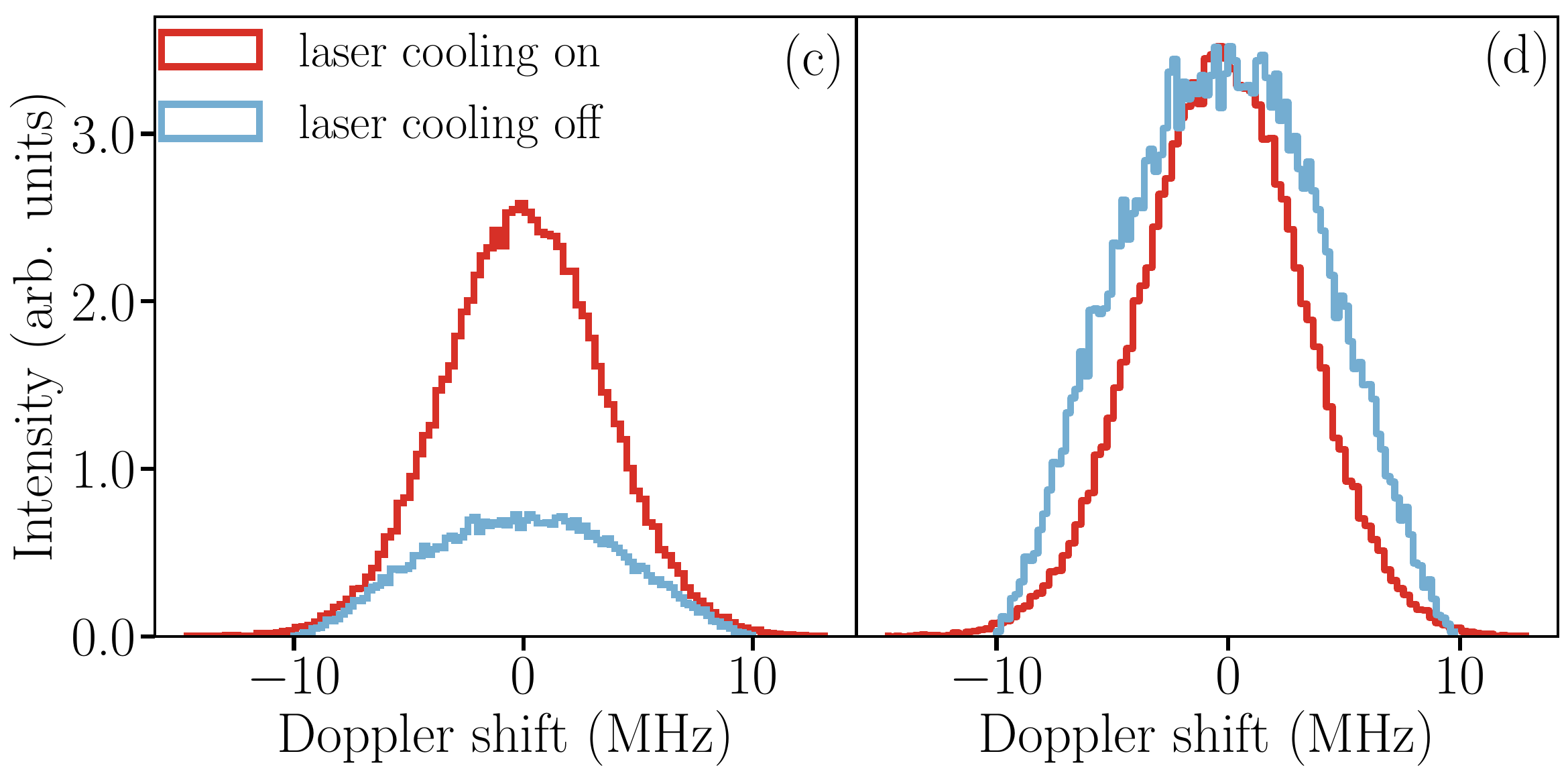}
\\
\includegraphics[width=0.83\columnwidth]{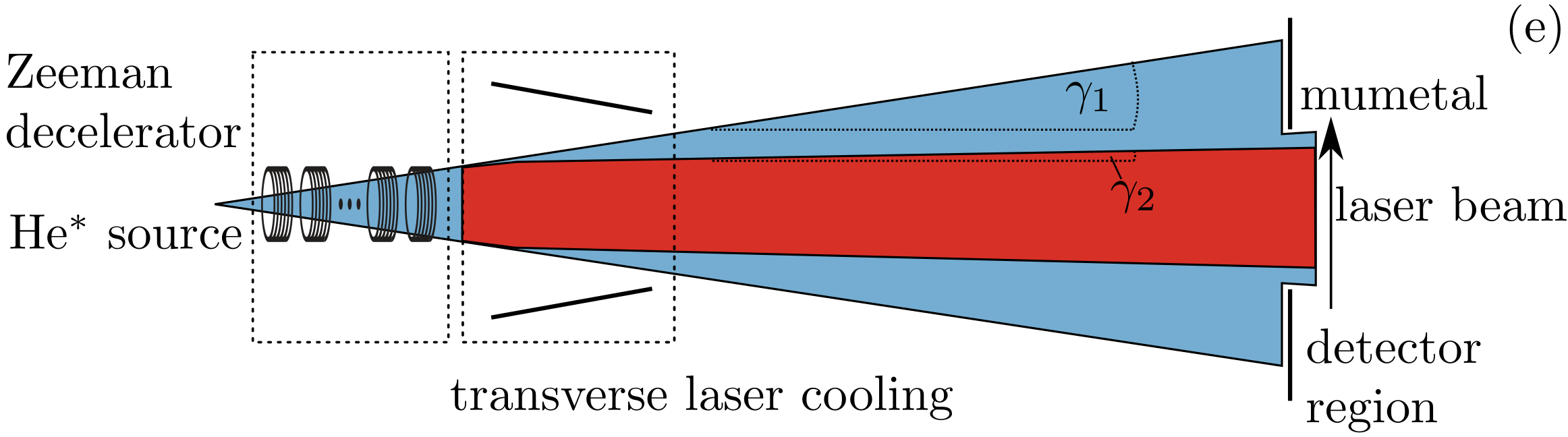}
\caption{\label{fig:6}  (a) Comparison of the spectrum of the $(1 s)(40 p) \,^3 P_J \, \leftarrow (1 s)(2s) \,^3 S_1$ transition of He$^*$ with the cooling laser turned on (red trace) and turned off (light-blue trace). Sticks (in black) represent the fine-structure components of the transition ($J=2,1,\,\mathrm{and}\;0)$. (b) Normalized spectra with laser cooling on (red trace) and off (light-blue trace). (c) Line profiles and intensities corresponding to the simulated velocity distributions after laser cooling (red trace) and without laser cooling (light-blue trace). (d) Same as (c), but after normalization to the same maximal intensity. (e): Illustration of the effect of the mumetal hole size on the atomic beam (light blue) and the laser-cooled beam (red). }
\end{figure}

 
To assess and optimize the efficiency of the transverse laser cooling, the transverse spatial distribution of the nondecelerated atomic cloud (i.e. without use of the Zeeman decelerator) is measured with the two on-axis imaging MCPs located beyond the laser-cooling section (see Fig. \ref{fig:1}). The effects of transverse laser cooling are clearly seen when comparing the images obtained with (Fig. \ref{fig:5}(c)) and without (Fig. \ref{fig:5}(d)) laser cooling, and when cooling in only one of the transverse ($x$ or $z$) directions (Figs. \ref{fig:5}(a) and \ref{fig:5}(b), respectively). The density profiles of the atomic cloud along the $x$ and $z$ directions at the positions of the imaging MCPs 1 and 2 are determined directly from the images, as illustrated in Figs. \ref{fig:5}(e) and \ref{fig:5}(f) for the $x$ direction, and follow a Gaussian distribution. Identical results (not shown) are obtained for the $z$ direction. The transverse-velocity distribution (see Fig. \ref{fig:5}(g)) is determined from the relative sizes of the He$^*$ cloud at the positions of the imaging MCPs 1 and 2, as explained in Sec. \ref{sec:measurement}. \\ \indent When the cooling lasers were turned off, the atomic beam expands from the outlet of the last coil of the Zeeman decelerator in a field-free region with expansion angle $\gamma_1$ (indicated in Fig. \ref{fig:6}(e) below). The FWHM of the atomic cloud increases in both directions from $1.07(2)\,\mathrm{cm}$ to $1.70(3)\,\mathrm{cm}$  over the distance of 38 cm separating the two detectors, corresponding to a mean transverse velocity of $|v_x|= 4.14(24)$ m/s. This result verifies the predicted transverse-velocity distribution of a nondecelerated atomic beam obtained by the Monte-Carlo particle-trajectory simulation (see Sec. \ref{sec:zeeman}). When the cooling lasers were turned on, the sizes of the atomic cloud monitored at imaging MCPs 1 and 2 were markedly reduced. The FWHMs of the images recorded at MCPs 1 and 2 (0.63(1) cm and 0.76(1) cm, respectively) imply a mean transverse velocity of $|v_x|=0.86(9)$ m/s, which corresponds to an expansion angle $\gamma_2\approx 1.7$ mrad (see Fig. \ref{fig:6}(e) below). This velocity is within a factor of 3 of the Doppler-cooling limit of the $2\,^3P_2\,\leftarrow\,2\,^3S_1$ transition \cite{chang14a}, which demonstrates the efficacy of the laser-cooling scheme.  \\ \indent The transverse-velocity distribution under the same conditions was also determined from the Doppler width of the spectra of the $(1s)(40 p) \,^3 P_J \, \leftarrow (1 s)(2 s) \,^3 S_1$ transitions, with the only difference that an 8-mm-diameter hole in the mumetal shield prevented the atoms with the largest transverse velocities from being detected. The results are displayed in Figs. \ref{fig:6}(a) and (b), which present the data before and after normalization to the same maximal intensity, respectively. 
To cancel the effect of the $1^\mathrm{st}$-order Doppler shift arising from a deviation $\alpha$  from $90^\circ$ of the angle between the laser and the supersonic beams, the laser beam is retroreflected after the photoexcitation region using a highly reflective mirror, which leads to a splitting of each transition into two components with opposite Doppler shifts $\pm \frac{v}{c}\nu_\mathrm{L} \sin \alpha $. The deviation angle $\alpha$ is chosen to be large enough to spectrally resolve the two Doppler components. Consequently, the transition appears as a pair of lines in Figs. \ref{fig:6}(a) and (b), and the $1^\mathrm{st}$-order-Doppler-free transition frequency can be determined as the average of the two Doppler-shifted frequencies \cite{hoelsch19a}. \\ \indent The fine-structure splittings of the $40p\,^3P_J$ ($J$=0--2) Rydberg state of He (from Ref. \onlinecite{drake99a} and indicated by the sticks in Figs. \ref{fig:6}(a) and (b)) are not resolved in the spectrum of the $(1 s)(40p) \,^3P_J \, \leftarrow (1s)(2s) \,^3S_1$ transition and do not contribute significantly to the line broadening. The width of each of the two Doppler-shifted components thus directly reflects the transverse-velocity distribution. Moreover, the relative intensities obtained from spectra recorded with and without laser cooling enable us to assess the increase in atom density in the photoexcitation region resulting from the laser cooling.\\ \indent   
The Doppler width of the $(1s)(40 p) \,^3 P_J \, \leftarrow (1s)(2s) \,^3 S_1$ transition observed when the cooling laser is turned off is 11.3(3) MHz (FWHM), corresponding to a mean transverse velocity of only 1.45(7) m/s, which is significantly less than the value of 3.8 m/s (corresponding to a Doppler width of $30\,\mathrm{MHz}$) estimated at the entrance of the laser-cooling section (see light-blue trace in Fig. \ref{fig:3}(b)). The reason for this discrepancy is that the 8-mm-diameter entrance hole in the mumetal shield prevents the atoms with largest transverse velocities from reaching the laser-excitation region, as illustrated schematically in Fig. \ref{fig:6}(e). When the cooling laser is turned on, the linewidth reduces to 7.9(2) MHz (FWHM) (see red trace in Fig. \ref{fig:6}(b)), corresponding to a mean transverse velocity of 1.02(5) m/s, in excellent agreement with the mean transverse velocity derived from the imaging experiments. Because of the strong collimation resulting from laser cooling, almost all atoms reach the laser-excitation region in this case, which explains the intensity increase by a factor of three observed in Fig. \ref{fig:6}(a).
\\ \indent 
Figures \ref{fig:6}(c) and (d) present spectra calculated from the velocity distribution in the photoexcitation volume of the detected He$^*$ atoms. These distributions were determined from particle-trajectory simulations, with initial conditions given by the light-blue trace in Fig. \ref{fig:3}, corresponding to the cooling laser being turned on (red trace) and off (blue trace). They include the effects of laser cooling and of the 8-mm-diameter hole in the mumetal shield and reproduce the experimental data well, in particular the factor-of-three increase of the signal intensity and the slight reduction of the linewidth in the spectrum recorded after laser cooling. One should note that the factor-of-three increase in signal strength corresponds to a factor-of-nine increase in density because the laser beam traverses the entire sample in $x$ direction and its diameter ($\approx$ 1 mm)  is much smaller than the atomic beam in $y$ direction and thus only probes a one-dimensional density distribution. \\ \indent 
These results demonstrate that the main advantage of transverse laser cooling in our experimental setup is to increase the atom density and thus the signal strength. Its effects on the Doppler width are largely offset by the rejection of the atoms with the largest transverse velocities by the hole of the mumetal shield in the experiments without laser cooling (see Fig. \ref{fig:6}(e)).
\\ \indent
\begin{figure}
\includegraphics[width=0.83\columnwidth]{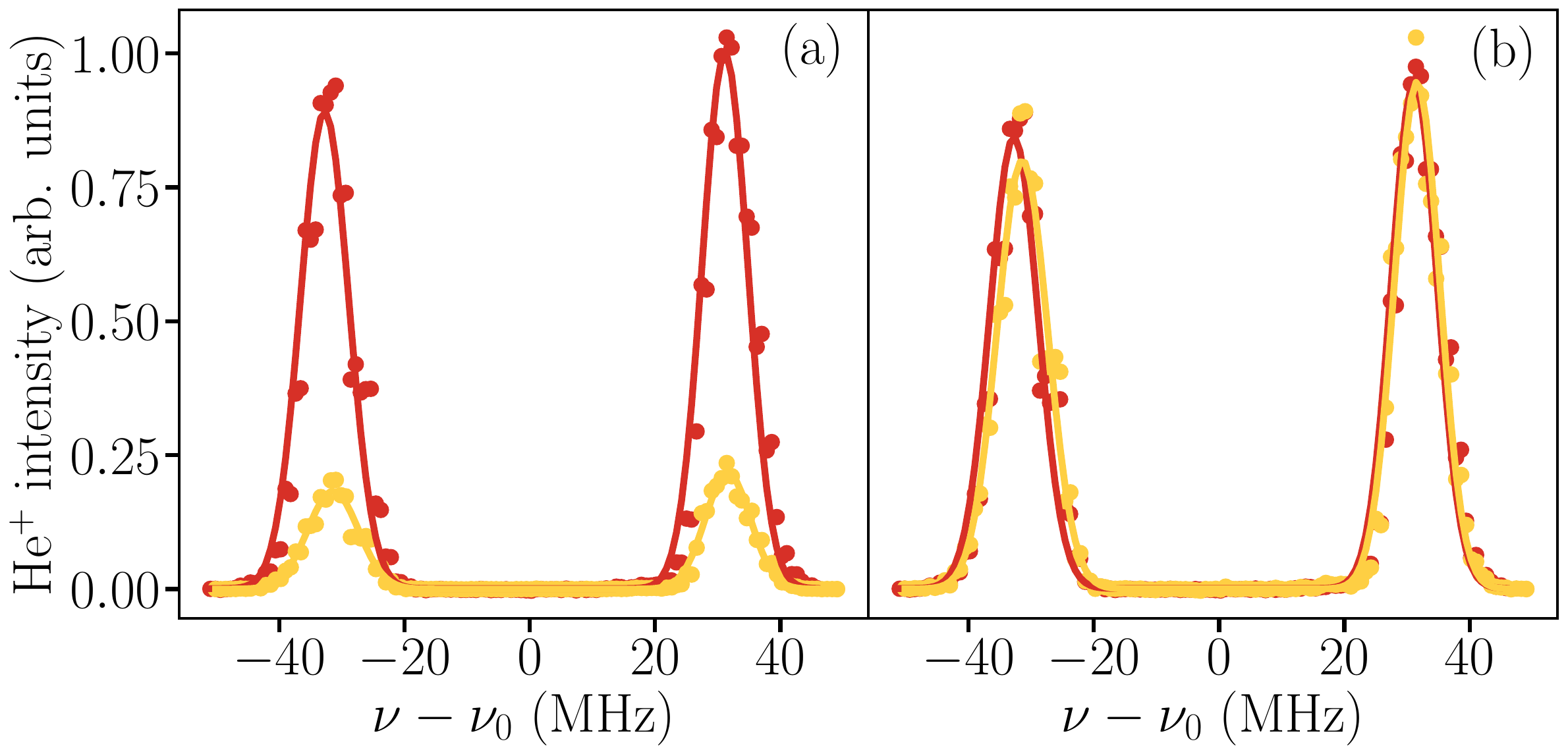} \\ \hspace{1.5 mm}
\includegraphics[width=0.83\columnwidth]{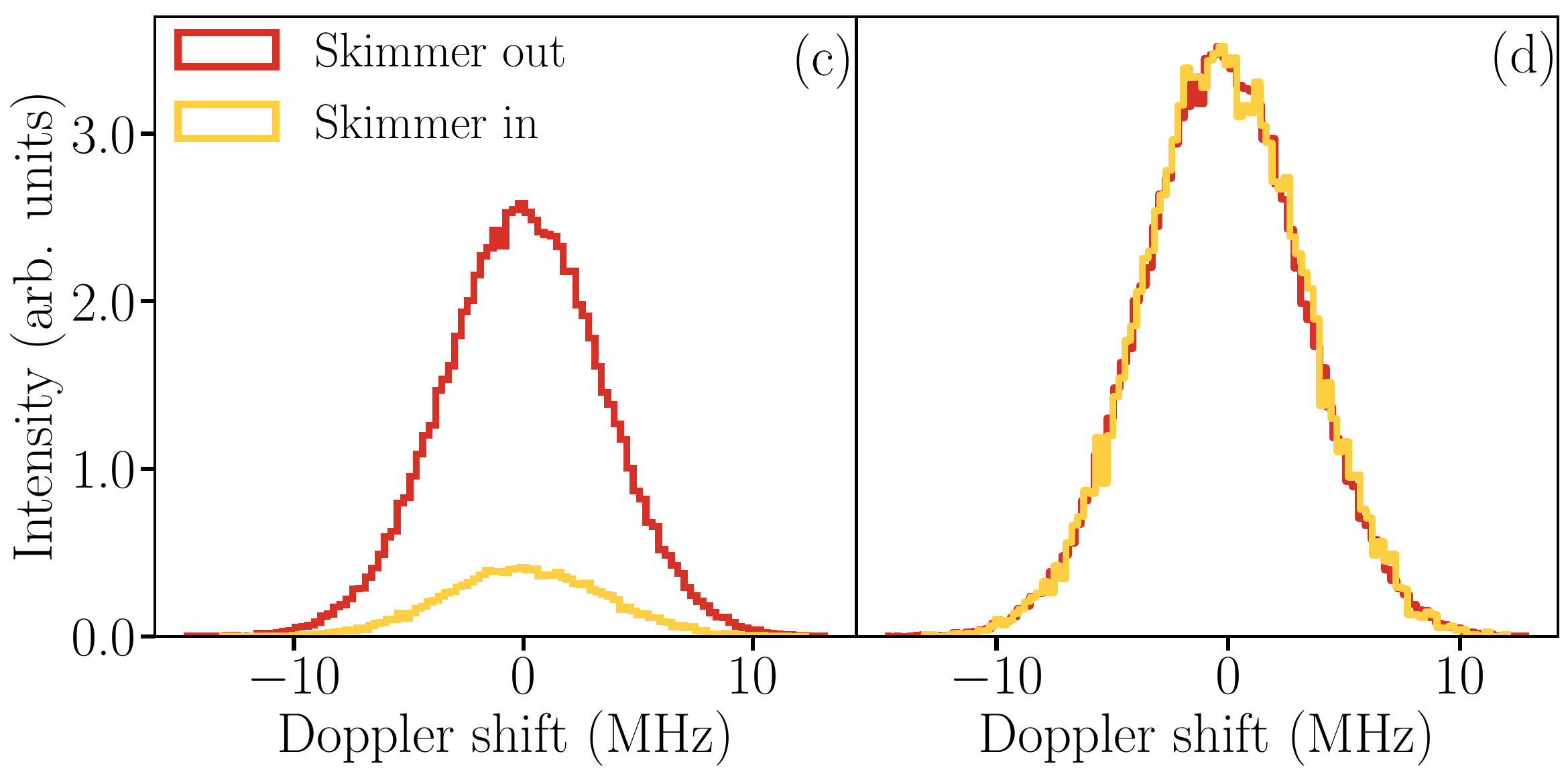} \\  \includegraphics[width=0.83\columnwidth]{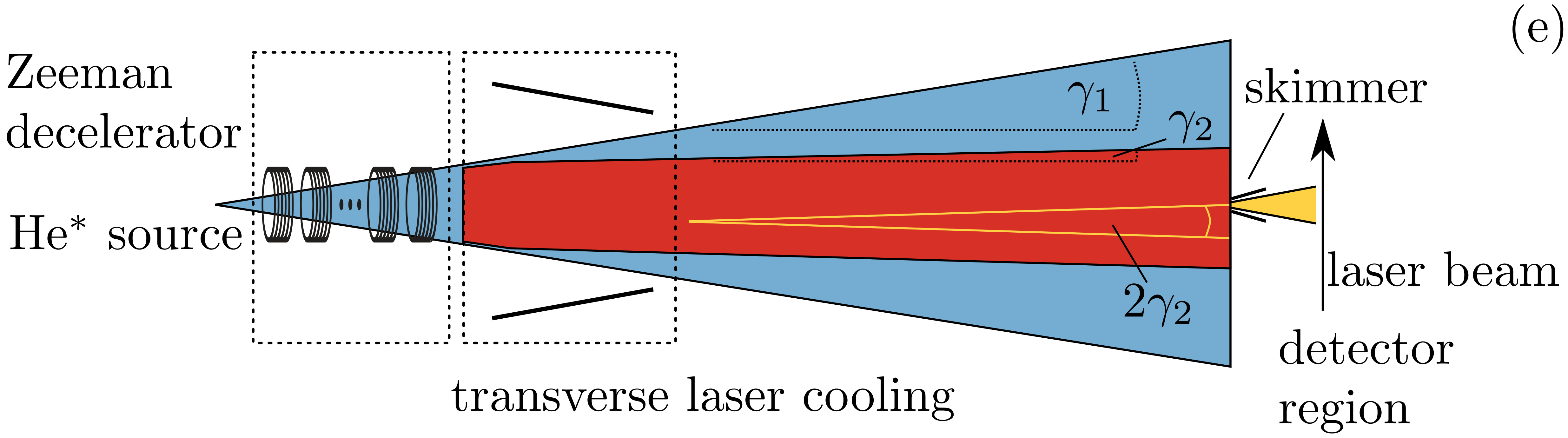} \caption{\label{fig:7}
(a) Comparison of the spectra of the $(1s)(40p) \,^3P_J \, \leftarrow (1s)(2 s) \,^3 S_1$ transition of He$^*$ in a laser-cooled beam with (yellow trace) and without (red trace) an additional skimmer. (b) Same as (a), but after normalization of the spectra to the same maximal intensity. (c) Spectral lineshape and intensities calculated using the velocity distributions of the laser-cooled He$^*$ beams determined from the numerical particle-trajectory simulations with (yellow trace) and without (red trace) additional skimmer. (d) Corresponding simulated Doppler linewidths with normalized intensities. (e) Schematic illustration of the trajectories of atoms corresponding to the results depicted in (a)--(d). } \end{figure}
Figure \ref{fig:7} demonstrates the effect of placing a skimmer or an aperture in front of the laser-excitation region on the intensity and the Doppler width of the $(1s)(40 p) \,^3P_J \, \leftarrow (1s)(2s) \,^3 S_1$ transition of He$^*$ after laser cooling. The red and yellow traces in panels (a) and (b) correspond to spectra recorded without skimmer and with a 1-mm-diameter skimmer placed 5 cm upstream from the photoexcitation spot, respectively. \\ \indent The main effect of the skimmer is to reduce the intensity by a factor of about 5 (see panel (a)). Comparing the two spectra after normalization to the same intensity (see panel (b)) reveals that the lineshapes are almost identical, with a FWHM of 7.9(2) MHz. Both effects are explained by the schematic diagram of the gas expansion in panel (e). When the cooling laser is turned off, the gas expansion of the atomic beam after the Zeeman decelerator corresponds to the light-blue cone with expansion angle $\gamma_1$. When the cooling laser is turned on, each point on the cross-sectional area of the He$^*$ beam at the exit surface of the laser-cooling section is the origin of a new expansion cone with expansion angle \begin{equation}\gamma_2= \arctan \frac{v_\perp}{v_\parallel}= 2.1 \,\mathrm{mrad}, \label{eq:gamma2} 
\end{equation}  given by $v_\perp \approx 1$ m/s and $v_\parallel=480$ m/s (see Sec. \ref{sec:results_cooling}). \\ \indent $\gamma_1$ and $\gamma_2$ are both very small and the excitation laser beam ($\lambda \sim 260$ nm) has a diameter comparable to the skimmer opening. Consequently, this laser beam acts as a slit along the $z$ direction and therefore the signal-intensity-reduction factor results from a one-dimensional selection and closely corresponds to the ratio of the laser-cooled-beam diameter at the skimmer orifice to the skimmer-orifice diameter itself (see Fig. \ref{fig:7}(e)). These considerations imply that the beam diameter at the end of the laser-cooling section is about 5 mm, which is in good agreement with the results obtained in Sec. \ref{sec:zeeman}. The reason why the linewidth is not reduced by the skimmer comes from the broad spatial distributions of He$^*$ at the end of the laser-cooling section: A simple geometric argument indeed demonstrates that atoms passing the skimmer can have a significant transverse velocity if they originate from an off-axis position at the exit plane of the laser-cooling section (see yellow trajectories in Fig. \ref{fig:7}(e).
\\ \indent The effects of the skimmer on the linewidths and line intensities can be reproduced quantitatively by propagating the atom trajectories through the laser-cooling section and considering the 8-mm-diameter hole in the mumetal shield, using as input the transverse-position distribution obtained from the Monte-Carlo simulations (see Fig. \ref{fig:3} in Sec. \ref{sec:zeeman}). Figure \ref{fig:7} (c) shows the distribution of Doppler shifts corresponding to the transverse velocities of the laser-cooled He$^*$ atoms that can pass through the mumetal hole (red trace) and through the 1-mm-diameter skimmer (yellow traces). Figure \ref{fig:7} (d) displays the corresponding normalized intensities. These results demonstrate that the transverse-velocity distribution and the Doppler width of a transversely laser-cooled sample cannot be significantly reduced by a skimmer or an aperture, at least as long as the condition \begin{equation} \label{eq:condition}
d_s \sin \gamma_2 \leq r
\end{equation} is fulfilled, where $d_s$ is the distance between the end of the laser-cooling section and the skimmer and $r$ is the radius of the atomic beam at the end of the laser-cooling section. To significantly reduce the Doppler width for $\gamma_2=2.1$ mrad, the skimmer would have to be placed more than $1.2\,\mathrm{m}$ away from the laser-cooling section. The same argument also implies that reducing the skimmer orifice diameter below $2r$ only leads to a signal loss without reduction of Doppler width. \\ \indent 
An alternative way of reducing the Doppler width with a skimmer is to reduce the beam velocity ($v_\parallel$ in Eq. \ref{eq:gamma2}) so that $d_s \sin \gamma_2$ in  Eq. 
(\ref{eq:condition}) becomes larger than $r$. In the next section, we show that a Zeeman decelerator can be used for this purpose prior to transverse laser cooling.
\subsection{Zeeman Deceleration}
The results of the particle-trajectory simulation presented in Sec. \ref{sec:zeeman} (see Fig. \ref{fig:3}(b)) indicate that Zeeman deceleration leads to a significant broadening of the transverse-velocity distribution. Decelerating the He$^*$ beam from 480 m/s to 175 m/s is accompanied by an increase of the FWHM of the transverse-velocity distributions by a factor of 2 (from $\pm 3.8\,\mathrm{m/s}$ to $\pm7.5\,\mathrm{m/s}$), corresponding to a FWHM of 58 MHz at the frequency of the $(1s)(40 p) \,^3P_J \, \leftarrow (1s)(2s) \,^3S_1$ transition. Moreover, this transverse velocity causes a large expansion angle of the supersonic beam, with $\gamma_2=42\,\mathrm{mrad}$ at 175 m/s. The estimated He$^*$ beam diameter in the photoexcitation zone 65 cm downstream from the decelerator is 5.4 cm. The He$^*$ density correspondingly decreases to the extent that it was not possible to detect the decelerated He$^*$ beam, neither with the imaging detectors nor by PFI following laser excitation. \begin{figure}[h]
\includegraphics[trim=0cm 0cm 0cm 0cm, clip=true, width=1.0\columnwidth]{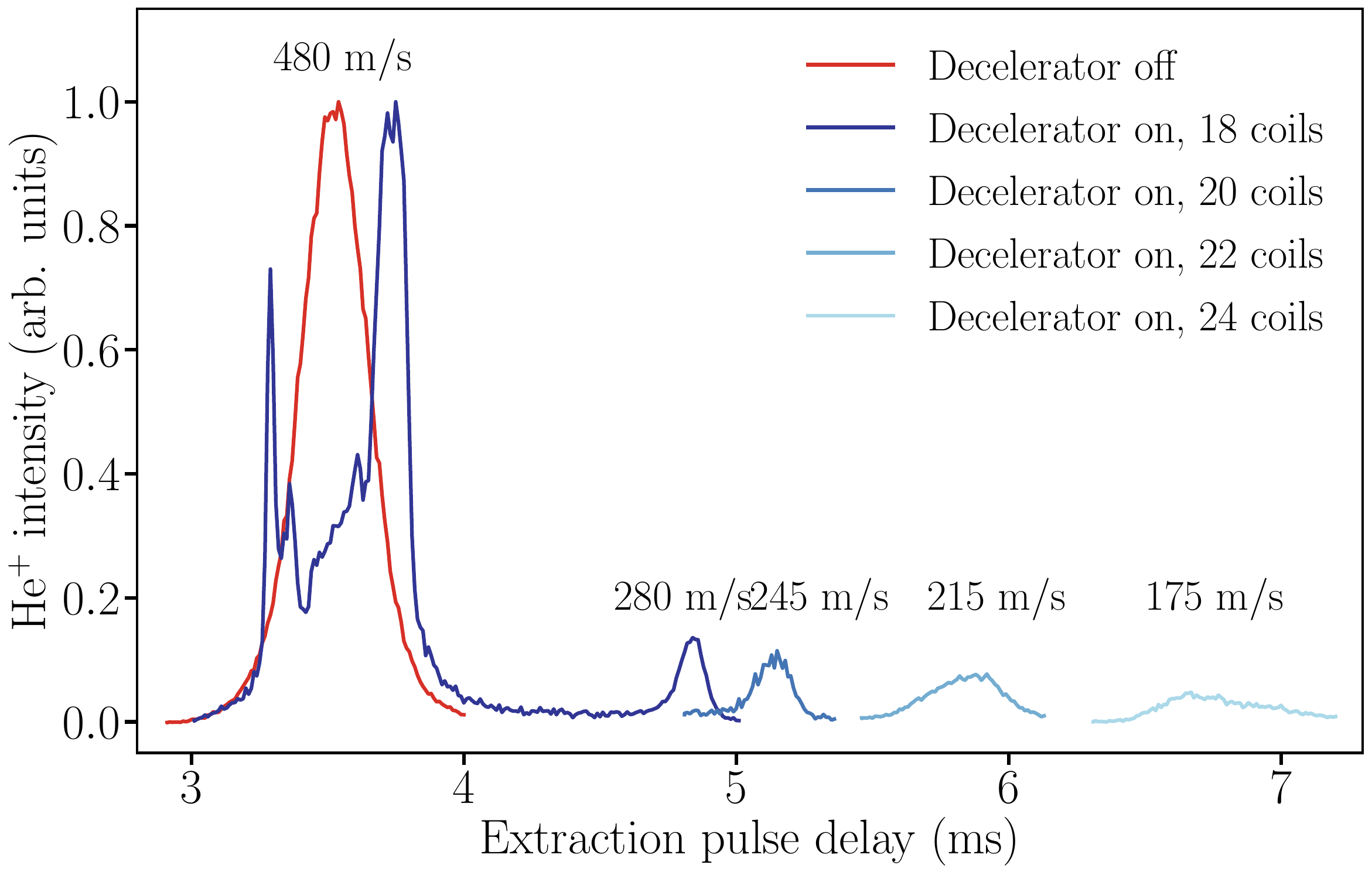}
\caption{\label{fig:8}Time-of-flight distributions of He$^*$ after Zeeman deceleration and transverse laser cooling measured by pulsed-field ionization following photoexcitation to the $(1 s)(40p) \,^3P_J$ Rydberg state. The red trace corresponds to the nondecelerated beam and the blue traces to beams decelerated using different numbers of coils, as indicated by the legend. }
\end{figure} 
\\ \indent The decelerated beam, however, was easily detected after laser cooling and Fig. \ref{fig:8} presents the resulting time-of-flight distributions of the He$^*$ pulses after deceleration to final velocities of 280, 245, 215, and 175 m/s (shown in shades of blue) using 18, 20, 22, and 24 coils in the deceleration sequence, respectively. For comparison, the temporal profile of the nondecelerated beam is displayed in red.  These measurements were carried out by monitoring the He$^+$ ion yield obtained by photoexcitation of the $(1s)(40p) \,^3 P_J \, \leftarrow (1s)(2s) \,^3S_1$ transition and PFI of the Rydberg states as a function of the extraction-pulse delay (see Sec. \ref{sec:measurement}). Operating the Zeeman decelerator leads to a pronounced modification of the pulse profile at short times of flight, as documented in earlier work \cite{wiederkehr10b,wiederkehr11a},
and to packets of slow, decelerated atoms that are detected at longer and longer times of flight as the number of coils in the deceleration sequence is increased. 
\begin{figure}[]
\includegraphics[trim=0cm 0cm 0cm 0cm, clip=true, width=1.0\columnwidth]{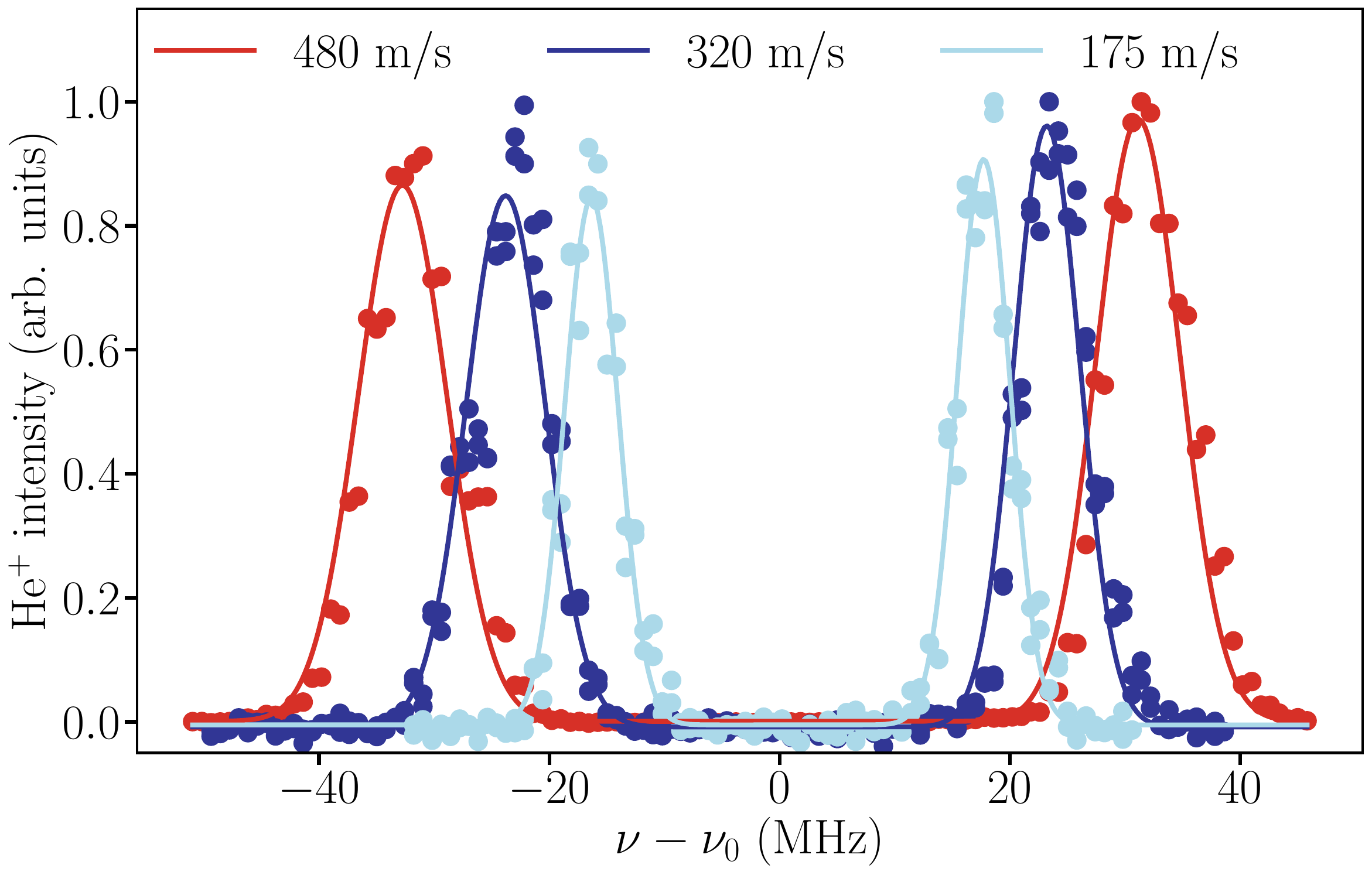}
\caption{\label{fig:9} Normalized spectra of the $(1s)(40p) \,^3P_J \, \leftarrow (1s)(2s) \,^3 S_1$ transition measured for a He$^*$ beam with a mean velocity of 480 m/s (red) and for He$^*$ beams decelerated to mean velocities of 320 m/s (blue) and 175 m/s (light blue) using a multistage Zeeman decelerator. }
\end{figure} 
\\ \indent 
Figure \ref{fig:9} compares the spectra of the $(1s)(40p) \,^3P_J \, \leftarrow (1s)(2s) \,^3S_1$ transition recorded after Zeeman deceleration to final velocities of 320 m/s (dark blue) and 175 m/s (light blue) with the spectrum obtained using a nondecelerated beam (red), after normalizing to the same maximal intensity. The splitting between the two Doppler components is proportional to the forward velocity of the atomic beam. The reduction in linewidth to 5.0(2) MHz in the spectrum recorded with the 175 m/s beam originates from the fact that for this slow beam only atoms with a transverse velocity of less than 0.65 m/s can pass through the hole in the mumetal shield surrounding the photoexcitation region. The high signal-to-noise ratio of this spectrum indicates that narrower linewidths could be achieved by increasing the flight distance from the Zeeman decelerator to the mumetal shield. As expected, the FWHM of the line of the spectrum recorded with the 320-m/s beam is intermediate between that obtained with the 175-m/s and the nondecelerated beams. 
\\ \indent This measurement illustrates the high-frequency resolution that can be achieved by combining Zeeman deceleration and laser cooling. At the signal-to-noise ratio of the spectrum recorded at 175 m/s, the central position of the line can be determined with a precision of $\frac{1}{100}$ of the linewidth, i.e., 50 kHz, which corresponds to an uncertainty of $\Delta \nu/\nu$ of $4\times 10^{-11}$.
\section{Conclusions}
Using the example of metastable He, we have demonstrated the combination of multistage Zeeman deceleration with transverse laser cooling to generate high-density supersonic beams with narrow transverse-velocity distributions. The Zeeman deceleration provided slow beams down to velocities of 175 m/s which are ideal for precision spectroscopy because of the prolonged transit times. Transverse laser cooling enabled the efficient reduction of the transverse-velocity distribution of the supersonic beam. Using the curved-wavefront approach of transverse laser cooling, a large capture velocity of 25 m/s could be reached, enabling the generation of intense beams with transverse velocities close to the Doppler limit. Such beams are of great potential for precision spectroscopy, in particular for Doppler-free two-photon spectroscopy. \\ \indent Combining experiment and numerical particle-trajectory simulations, we have explored a broad range of experimental conditions such as atomic beam- velocities and densities, and geometrical constraints caused by skimmers and apertures. In the process, we have characterized and optimized the beam properties using imaging and time-of-flight methods in combination with spectroscopic measurements of the Doppler widths. The factors limiting the signal strength and resolution that can be achieved in single-photon spectroscopic experiments have been established. 
\\ \indent We have also demonstrated the power of the combination of Zeeman deceleration and transverse laser cooling in single-photon spectroscopic measurements with the example of the $(1s)(40p) \,^3P_J \, \leftarrow (1 s)(2s) \,^3S_1$ transition. In particular, we have obtained spectral lines with full-widths at half maximum as narrow as 5 MHz at UV frequencies around $1.15\times 10^{15}$ Hz, corresponding to $\Gamma_\mathrm{FWHM}/\nu$ of $4\times 10^{-9}$, with statistics sufficient to extract line centers at $\Delta \nu/\nu $ of $4\times 10^{-11}$ which is excellent for spectroscopic investigations in the UV frequency range. Additionally, we have elucidated how the linewidths might be further reduced by extension of the flight distances. The long transit times resulting from Zeeman deceleration and the high densities and narrow Doppler widths resulting from transverse laser cooling offer ideal conditions for precision spectroscopy in paramagnetic atoms and molecules amenable to laser cooling.

\begin{acknowledgments}
We thank Prof. Ben Ohayon for useful discussions on the curved-wavefront approach. We thank Anna Hambitzer, Lukas Gerster, Max Melchner, Dominik Friese, Lukas Möller, Marie-Therese Phillip, Sebastien Garcia and Matthias Stammeier for the contributions to the initial setup and
characterization of the apparatus. This work is supported financially by the Swiss National Science Foundation under grants No. CRSII5-183579 and No. 200020B-200478.
\end{acknowledgments}

%

\end{document}